\documentclass[twocolumn,letter]{jpsj2} 

\def\nle{\ \raise.3ex\hbox{$<$}\kern-0.8em\lower.7ex\hbox{$\sim$}\ }
\def\nge{\ \raise.3ex\hbox{$>$}\kern-0.8em\lower.7ex\hbox{$\sim$}\ }
\def\lcrh{L_h}
\def\Tc{T_{\rm c}}
\def\Tcsp{T^*}
\def\Tg{T_{\rm g}}
\def\Tirr{T_{\rm irr}}
\def\Tstp{T_{\rm stp}}
\def\FeMnTiO{Fe$_{0.5}$Mn$_{0.5}$TiO$_3$}
\def\FeMnTiOII{Fe$_{0.55}$Mn$_{0.45}$TiO$_3$}

\def\runauthor{Petra E. \textsc{J\"onsson}$^{1}$ and Hajime \textsc{Takayama}$^{2}$ }

\title{``Glassy Dynamics'' in Ising Spin Glasses --- Experiment and Simulation }
\author{ Petra Erika \textsc{J\"onsson}\thanks{Present address: RIKEN, Hirosawa 2-1, Wako, Saitama 351-0198; E-mail: pjonsson@riken.jp} 
and Hajime \textsc{Takayama}\thanks{E-mail: takayama@issp.u-tokyo.ac.jp} }
\inst{Institute for Solid State Physics, University of Tokyo,
5-1-5 Kashiwa-no-ha, Kashiwa, Chiba 277-8581}
\recdate{\today}
\abst{
The field-cooled magnetization (FCM) processes of Ising spin glasses under
relatively small fields are investigated by experiment on
\FeMnTiOII\ and by numerical simulation on the three-dimensional
Edwards-Anderson model. Both results are explained in a unified manner by
means of the droplet picture. In particular, the
cusp-like behavior of the FCM is interpreted as evidence, not for an
equilibrium phase transition under a finite magnetic field, but for a
dynamical (`blocking') transition frequently observed 
in glassy systems.
} 

\kword{Ising spin glass, field-cooled magnetization, droplet picture,
glassy dynamics}

\begin{document}
\sloppy
\maketitle

Slow relaxational dynamics of nonexponential type has been observed in   
various disordered systems, and is now called `glassy
dynamics'~\cite{rev-glassyDyn}. The most interesting slow dynamics is the
one observed in systems, such as structural glasses and spin glasses,
whose relaxation processes are determined cooperatively through
interactions between constituent elements of the system. In this case,
the distribution of relaxation times exhibits a peculiar temperature
dependence besides the one attributed to the thermally activated
process.  A typical example is the critical slowing down in Ising spin
glasses, which are considered to possess an equilibrium phase transition
at a finite temperature, $\Tc$, under zero magnetic
field.  On the other hand, a glass transition temperature, $\Tg$, of a
structural glass is considered to be a `blocking' temperature in the
sense that relaxation modes, whose relaxation time significantly exceeds
the experimental time window, appears to be frozen below
$\Tg$. By the word `glassy dynamics', here, we mean such slow dynamics
of a cooperative 
origin which, at the same time, involves thermal `blocking' at low
temperatures. We argue in the present letter that the
field-cooled-magnetization (FCM) process in Ising spin glasses is one 
typical example of such `glassy dynamics'.  

In spin-glass (SG) studies, one of most notable problems yet unsettled
concerns the stability of the equilibrium SG phase under a finite
magnetic field $h$~\cite{review1}. This is the case even for Ising spin
glasses which are the conceptually most simple SG systems. The
mean-field theory predicts the stability of the SG phase up to a certain
critical magnitude of $h$ specified by the de Almaida-Thouless (AT)
line~\cite{AT-line}, while according to the droplet
theory,~\cite{FH-88-NE,FH-88-EQ,BM-chaos} the SG phase is unstable even
in an infinitesimally small $h$. Recently, Takayama and Hukushima
(TH)~\cite{TH} have carried out numerical simulations of the field-shift
aging protocol on the three-dimensional (3D) Ising Edwards-Anderson (EA)
model, the results of which strongly support the droplet picture. They
have claimed that the AT-transition-like behavior observed
experimentally~\cite{KatoriIto94} is well interpreted as a dynamical
crossover from the SG behavior to the paramagnetic one upon application
of a field $h$.~\cite{Mattsson95} Further numerical works in favor of
the droplet picture have been reported more 
recently.~\cite{Y-Katz, Krza} 

Since early SG studies,\cite{Nagata} it has been observed
that the zero-FCM (ZFCM) deviates from the FCM at a certain temperature,
denoted here as $\Tirr$, and that the FCM exhibits a cusp-like behavior
at a certain temperature, denoted here as $\Tcsp$, below 
which the FCM becomes nearly independent of $T$. Here, the ZFCM (and FHM
used below) is the magnetization measured in a reheating process after a
ZFC (FC) process. Without consideration of cooling/heating rate effects,
these phenomena have often been considered to be evidence for the SG
phase transition at $\Tirr \simeq \Tcsp$ under a finite field.
The main purpose of the present study is to address this problem by
performing detailed experiments on the Ising spin glass \FeMnTiOII\ with
$\Tc \simeq 22.3$~K~\cite{our-ac}, by extending the Monte Carlo
simulation on the same Gaussian Ising EA model as in the study of TH
with $\Tc \simeq 0.95$~\cite{Marinari-cd98-PS,MariCamp} (in units
of the variance of interactions), and by comparing the results of the
two analyses. The consequence is that the above-mentioned FCM behavior,
a most fundamental phenomenon in spin glasses, can in fact be
interpreted as the `glassy dynamics' in the sense described  
above. 

\begin{figure}
\begin{center}
\resizebox{0.42\textwidth}{!}{\includegraphics{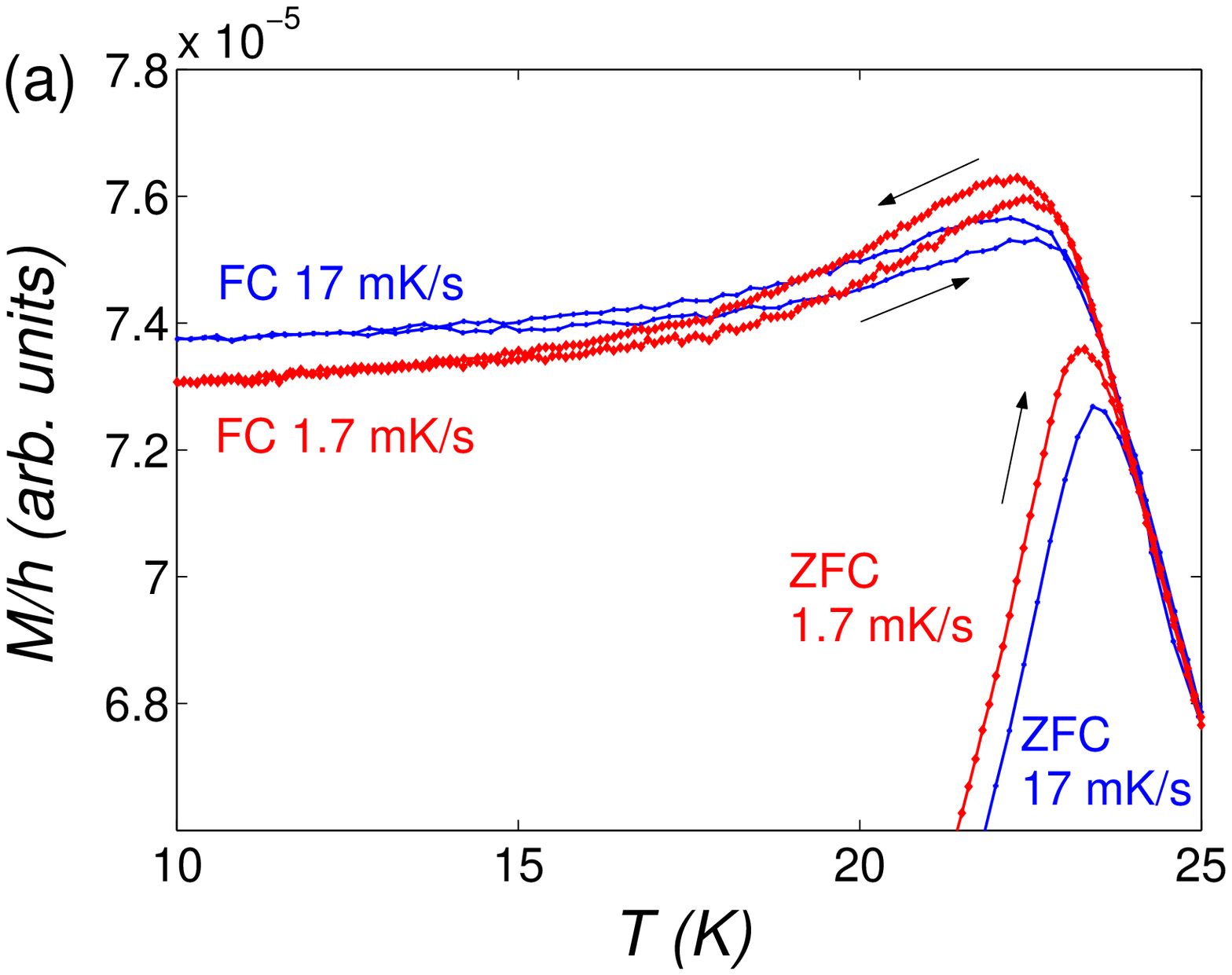}}
\hspace*{5mm}\resizebox{0.41\textwidth}{!}{\includegraphics{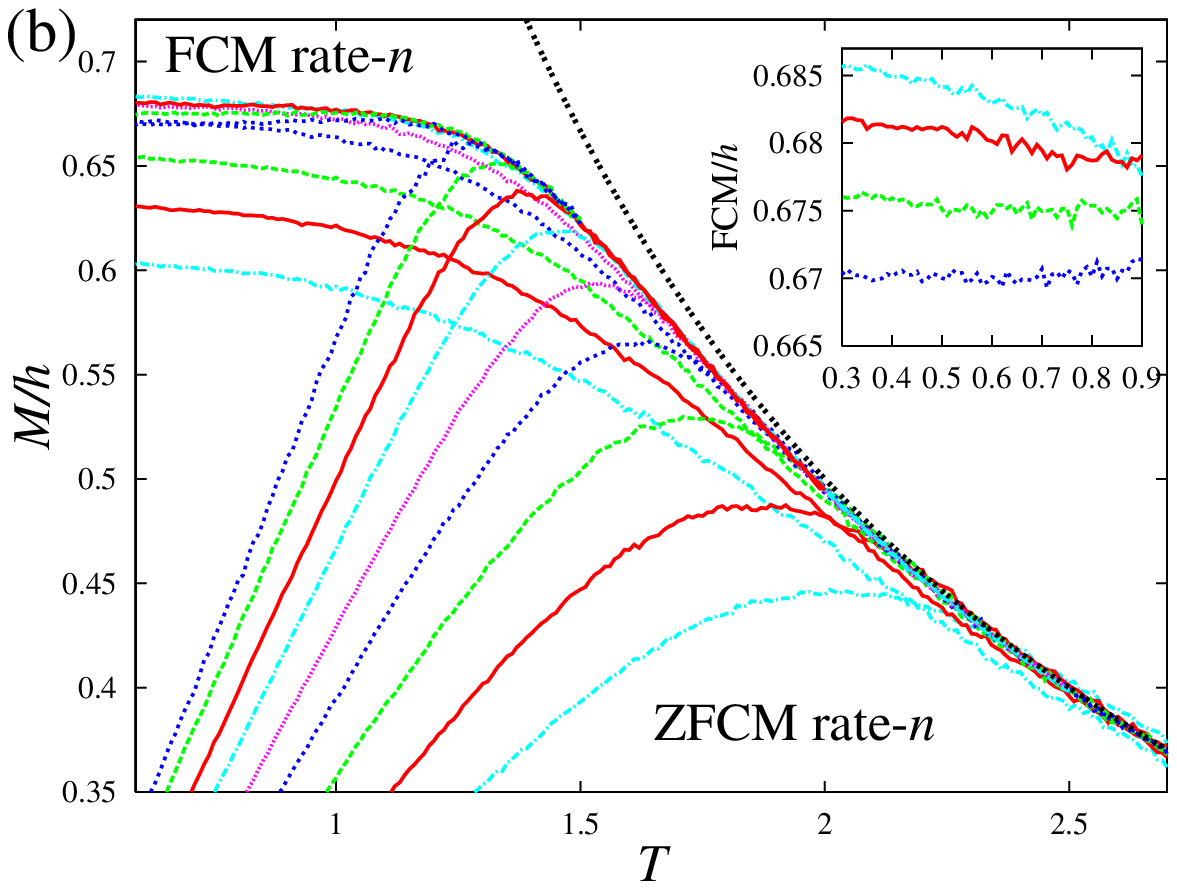}}
\end{center}
\vspace*{-3mm}
\caption{Cooling/heating rate dependences of FCM, FHM and ZFCM of (a)
 \FeMnTiOII\ with $h=5$~Oe, and (b) EA  model with $h=0.1$ and
 rate-1, 3, 10, 33, ... 3333, and 10000 (without  FHM). The arrows in
 (a) indicate the cooling/heating measurements. The line in (b)
 represents the Curie law. In the inset of (b) the enlarged FCMs with rate-333,
 1000, 3333 and 10000 (from top) are shown. The averages were taken
 over at least 3800 samples for the data used in the present work.}  
\label{fig:FCM-ZFCM}
\end{figure}

Let us begin with two sets of FCM, FHM and ZFCM curves, measured with two different cooling(heating) rates,
and shown in Fig.~\ref{fig:FCM-ZFCM}(a). It is clear
that $\Tirr$, the temperature at which the irreversibility between the
FCM and ZFCM occurs, does depend on the cooling rate. Furthermore, 
we can see that $\Tirr$ is significantly higher than $\Tcsp$, the
cusp-like temperature of the FCM.\cite{foot-Tstar}
The corresponding FCM and ZFCM obtained in our simulation are presented
in Fig.~\ref{fig:FCM-ZFCM}(b). The rate-$n$ in the figure means the
cooling process with $n$ MC steps (mcs) at each $T$, which is changed in
steps of $\Delta T =0.01$. When the rates are measured in units of
$\Tc/t_0$, with $t_0$ being the microscopic spin flip time (which is 1~mcs
for simulation and $10^{-12}$~s for experiment), the rate of the process
with rate-100 in the simulation is $1\cdot10^{-4}$, which has to be compared,
for example, with $2\cdot10^{-17}$ of that in the experimental process
denoted as 1.7~mK/s. The results in Fig.~\ref{fig:FCM-ZFCM} imply that the
occurrence of the irreversibility between the FCM and ZFCM corresponds
by no means to a certain equilibrium phase transition. 
  
\begin{figure}
\begin{center}
\resizebox{0.41\textwidth}{!}{\includegraphics{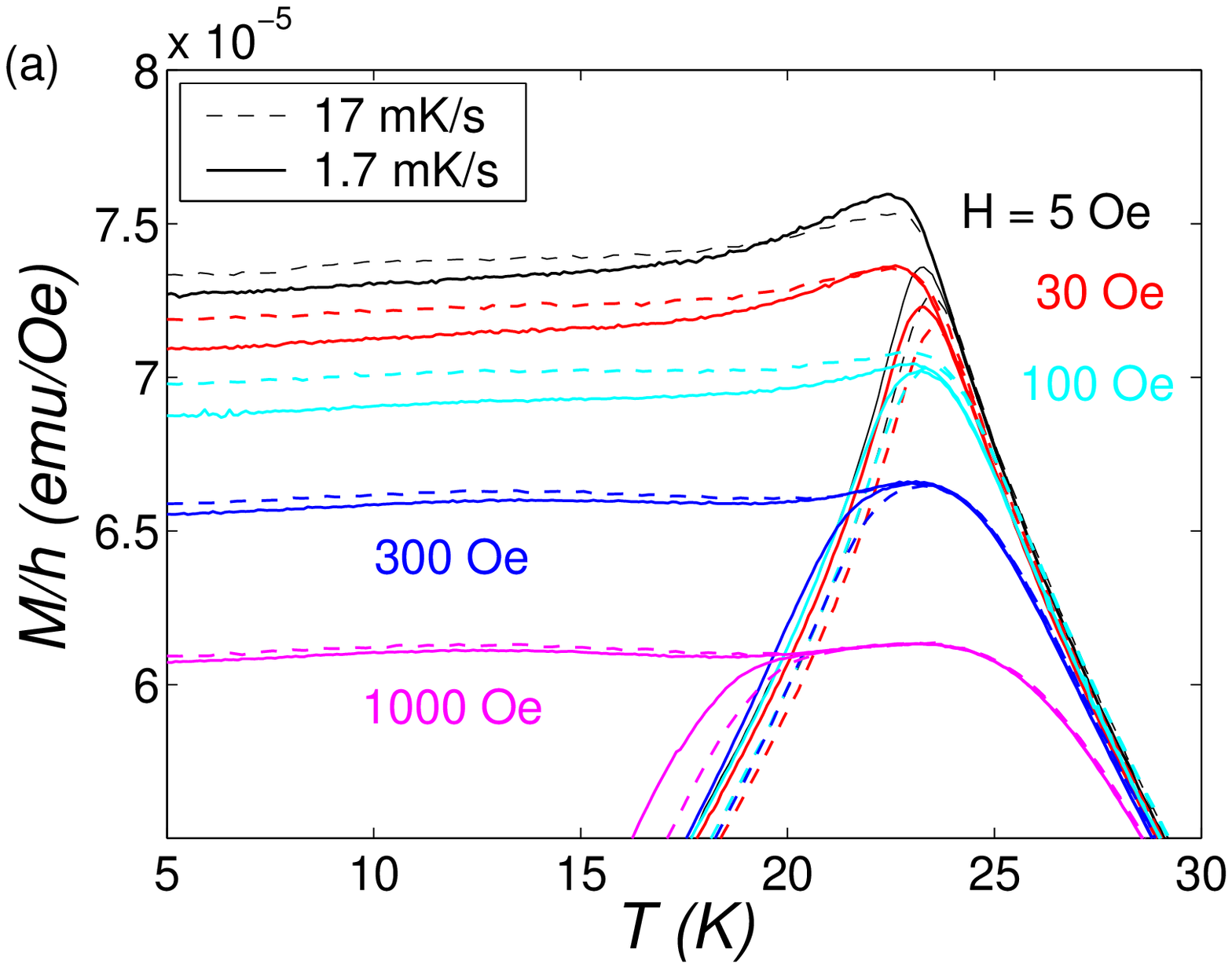}}
\hspace*{3mm}\resizebox{0.41\textwidth}{!}{\includegraphics{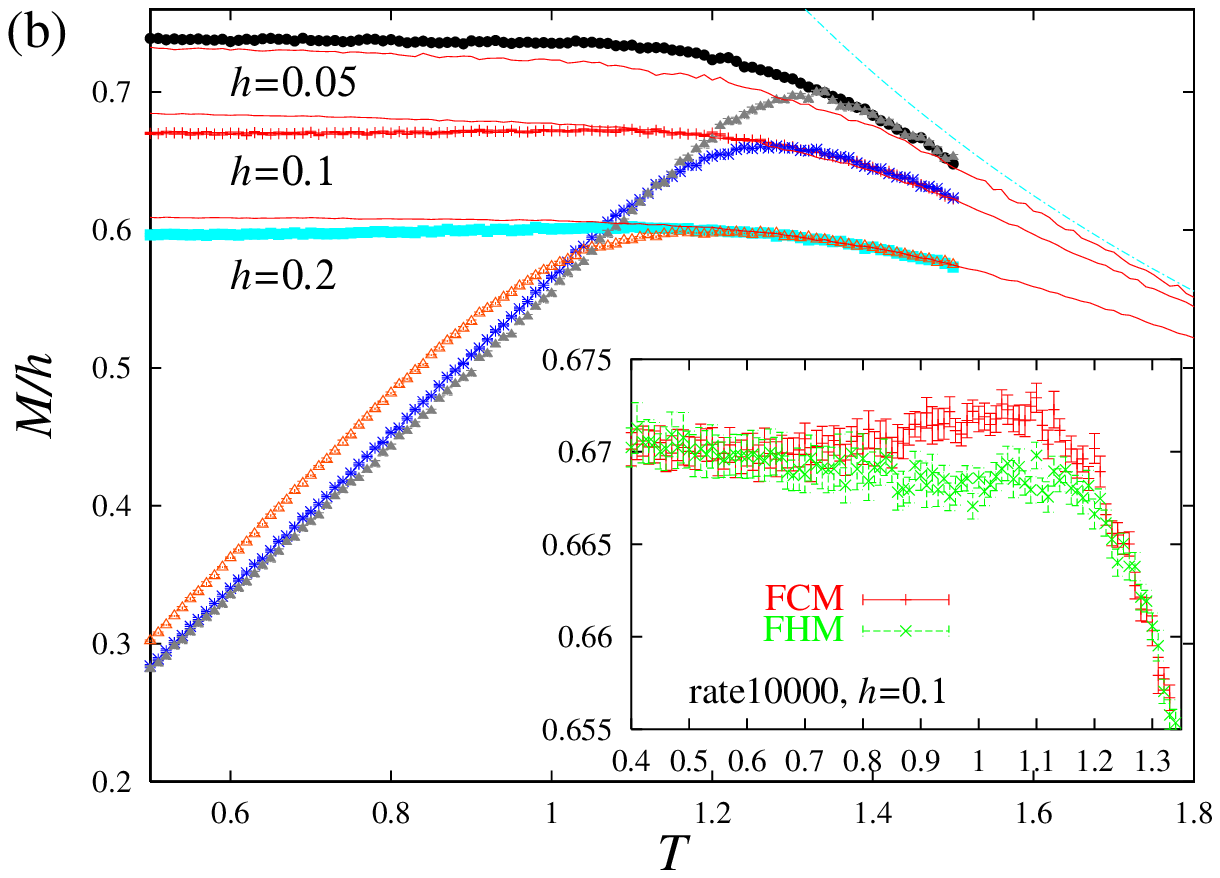}}
\end{center}
\vspace*{-3mm}
\caption{FCM and ZFCM curves under different fields and with
 different cooling rates for \FeMnTiOII\ (a), and EA model with
 rate-10000 (b). In (b) the corresponding FCM curves with rate-333 are
 shown by thin lines and in the inset the FCM and FHM curves are shown
 at an enlarged scale.
}
\label{fig:fcm-zfcm_h}
\end{figure}

At $T>\Tirr$, the FCM, which coincides with the ZFCM, is the
magnetization $M$ in equilibrium under $h>0$. The small deviation of
this equilibrium $M/h$ from the Curie law may be attributed to the
nonlinear field effect in the simulation with $h=0.1$ (see
Fig.~\ref{fig:fcm-zfcm_h}(b) below), and to a small nonvanishing
Curie-Weiss constant for the experiment with $h=5$~Oe. Here, we note that 
$h=0.1$ in the simulation corresponds to the order of 1~T in the
experiment on \FeMnTiOII. In this temperature range, the SG short-range
order, represented by the SG coherence length $\xi(T)$, and the
corresponding correlation time,  $\tau(T)$, increase, and at 
$T \simeq \Tirr$ the latter reaches the time scale proportional to the
inverse of the cooling rate. Actually, if this proportionality constant
is set to unity, and $\Tirr$ is estimated using the peak temperature of the
ZFCM, our simulation results can be fitted to an expression of the
critical slowing down 
\begin{equation}
\tau(T) \simeq a_0[(T-\Tg(h))/\Tg(h)]^{-z\nu},
\label{eq:c-tau}
\end{equation}
with $a_0 \simeq 30, z\nu=11.0$ and $\Tg(h=0.1)\simeq 0.81$, where 
$\Tg(h)$ is a supposed transition temperature under a finite $h$. Here, 
we don't claim that $\Tg(h)$ is an equilibrium transition temperature.
Instead, we want to emphasize that the SG short-range order starts to
increase at temperatures as high as $2\Tc$, with the value of $z\nu$
comparable to that obtained for the equilibrium transition under
$h=0$~\cite{our-ac, Gunnarsson}. We therefore consider that the SG
correlation of the order of $\xi^* \sim \tau^{1/z}(\Tirr)$ has appeared
at $T \simeq \Tirr$ on average, and that locally correlated
regions of relatively large size start to be thermally blocked
around $\Tirr$. We call them spin clusters.
   
\begin{figure}
\begin{center}
\resizebox{0.42\textwidth}{!}{\includegraphics{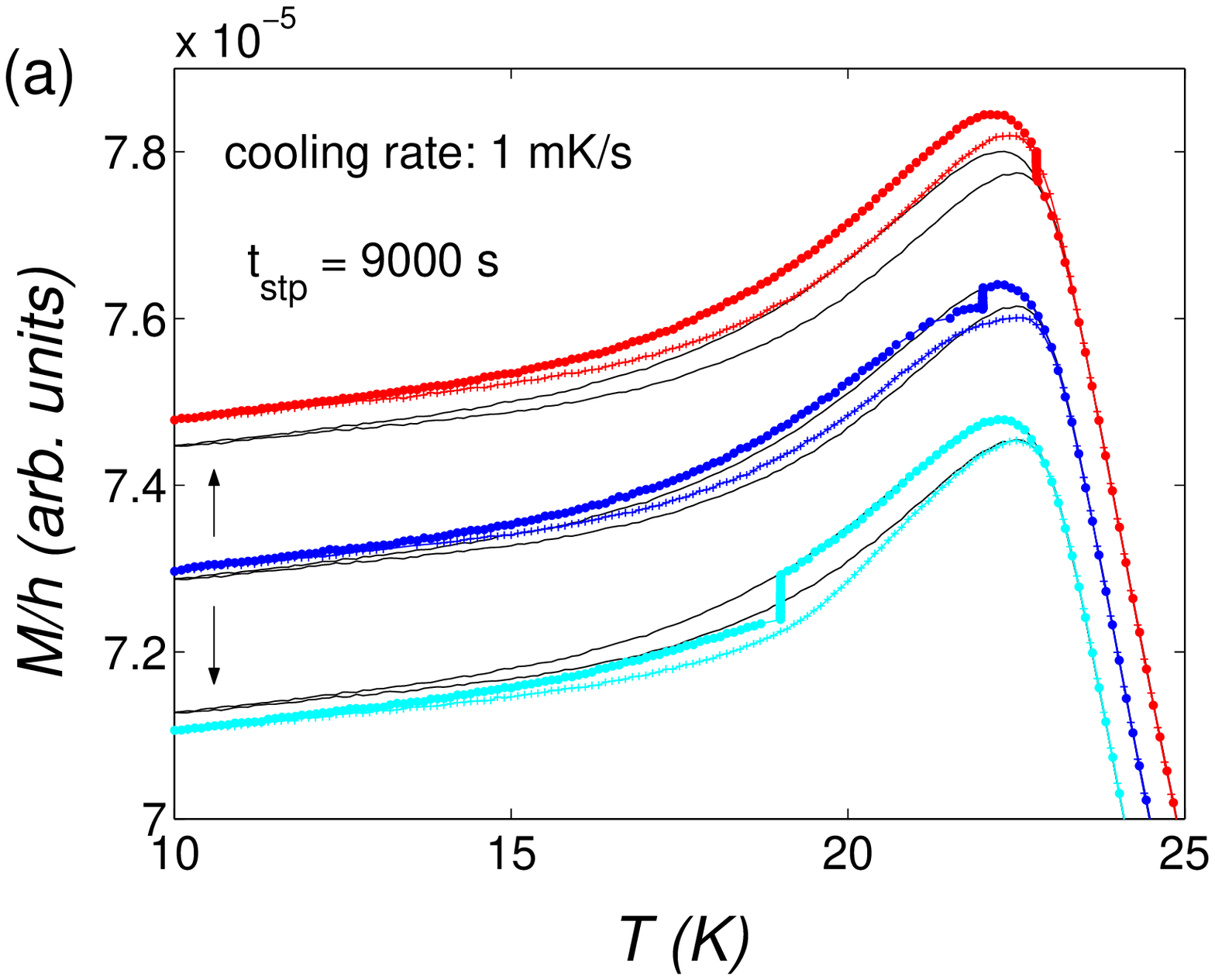}}
\resizebox{0.44\textwidth}{!}{\includegraphics{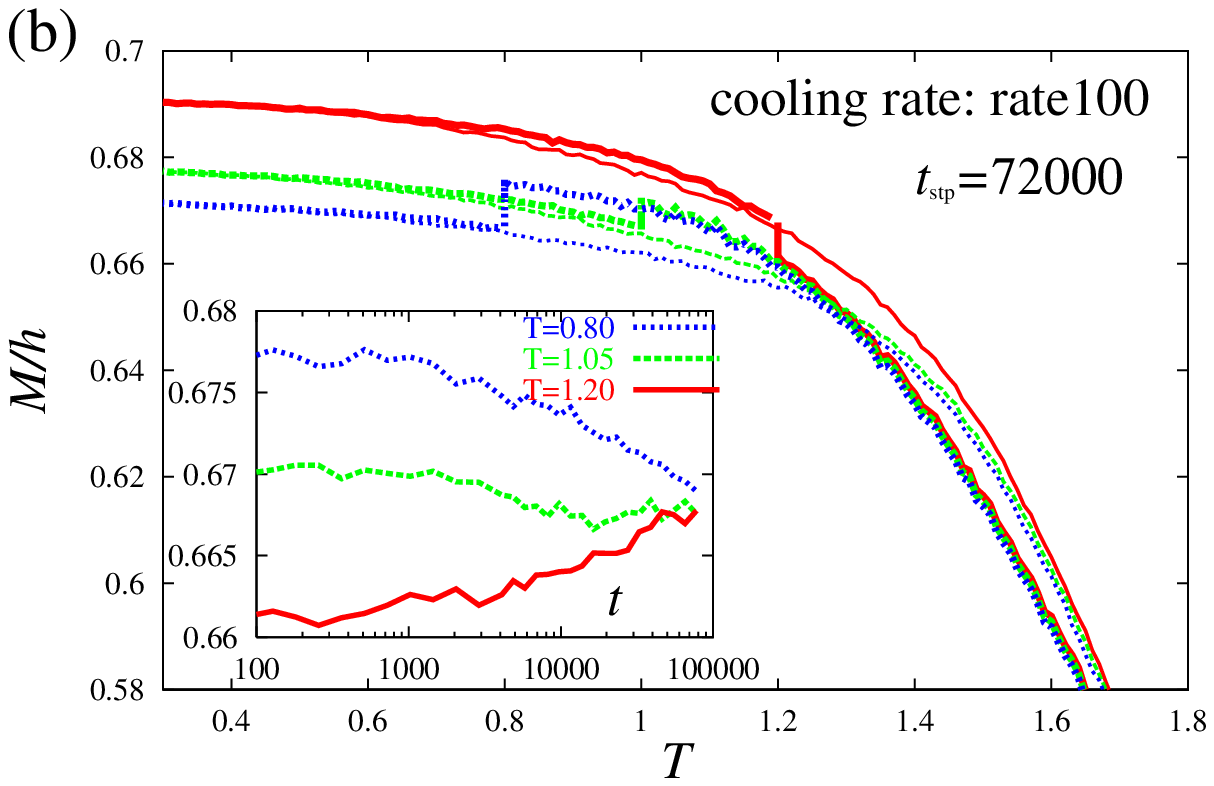}}
\end{center}
\vspace*{-3mm}
\caption{The FCMs (FHMs) observed in processes with a stop
 of the cooling at $\Tstp$ for a period $t_{\rm stp}$ are shown by thick
 solid (dashed) curves ($h=5$~Oe and 0.1). (a) The results for
 \FeMnTiOII\ with  
 $\Tstp=$22.8~K, 22~K and 19~K (top to bottom). Here, the upper and
 lower curves are shifted in height for clarity, and the FCMs (FHMs)
 without the halt of cooling are also shown by thin solid (dashed)
 curves. The  inset of (b) shows the time evolution of FCM during the stop. 
}
\label{fig:cl-T-cl-ht}
\end{figure}

At $T < \Tirr$, not only the ZFC state but also the FC state is out of
equilibrium. This is clearly seen in FCM processes, where the cooling is stopped at a certain temperature, denoted as $\Tstp$. As shown 
in Fig.~\ref{fig:cl-T-cl-ht}, the FCM increases during a stop made in the temperature range
$\Tirr > \Tstp > \Tc$ ($\Tstp=22.8$~K in the experiment and $\Tstp=1.2$
in the simulation). Because of a fairly large dynamical exponent $z$ in
eq.~(\ref{eq:c-tau}), the size of the SG short-range order is not
expected to appreciably increase further during the stop 
within this temperature range. Instead, spin clusters thermally
blocked at around $\Tirr$ tend to equilibrate, or further
polarize, yielding the FCM increase observed both in the experimental
and simulation results. In an FC process without such a stop, spin
clusters of a smaller size than the already blocked spin clusters are
blocked, and at around $\Tcsp$, these blocked clusters are
considered to come into contact with each other. 

The observed FCM behavior at temperatures comparable to and lower
than $T^*$ can be interpreted in terms of the droplet 
picture~\cite{FH-88-NE,FH-88-EQ,BM-chaos}, where the {\it field
crossover (correlation) length}~\cite{FH-88-EQ}, $\lcrh$, plays
important roles. $\lcrh$ separates the characteristic behavior of
droplet excitations in the equilibrium SG phase below $\Tc$ by their
size $L$: it is dominated by the Zeeman energy 
($\sim \sqrt{q_{\rm EA}}hL^{d/2}$) for $L>\lcrh$ and by the SG stiffness 
energy ($\sim \Upsilon L^\theta$) for $L<\lcrh$. Here, $q_{\rm EA}$ is
the SG order parameter, $d$ the spatial dimension, $\Upsilon(T)$ the
stiffness constant of SG domain walls, and $\theta$ the gap
exponent. Explicitly, $\lcrh$ is written as
\begin{equation}
\lcrh \sim (\Upsilon(T)/h\sqrt{q_{\rm EA}})^{\delta}
      \sim [((\Tc-T)/\Tc)^{a_{\rm eff}}/h]^{\delta}, 
\label{eq:lcrh}
\end{equation}
where $\delta = (d/2 - \theta)^{-1}$ and the exponent 
$a_{\rm eff}\ (>0)$~\cite{TH,our-ac} is given by the temperature
dependence of $\Upsilon$ and $q_{\rm EA}$. 

We now apply this droplet picture to spin clusters grown in the 
nonequilibrium FC process. At temperatures higher than  
$\Tcsp (\approx \Tc)$, spin clusters are thermally blocked
independently of each other, or in other words, the SG stiffness energy
introduced above is not effective at all. Thus, the configurations of the
spin clusters are overbalanced to the Zeeman energy. At and below $T^*$,
the SG stiffness energy becomes effective and competes with the
overbalanced Zeeman energy. We then naturally expect 
reorientation of spins, which results in the growth of new spin
clusters with SG short-range order for a given set of $(T,h)$ within
old spin clusters.~\cite{our-Suppl} 
We denote the mean size of the new spin clusters
simply as $\xi$. The consequences of this restoration of the SG
stiffness energy, or the growth of $\xi$, are not only the FCM cusp-like
behavior at $\Tcsp$ but also a decrease in FCM by further
cooling the system below $\Tcsp$. For the same reason, an {\it initial
decrease} in FCM is expected during a stop of the FCM process at
$\Tstp$ below $\Tcsp$.  All of these expectations were in fact observed
in the experiment as can be seen in Figs.~\ref{fig:FCM-ZFCM}(a) and
\ref{fig:cl-T-cl-ht}(a).  In the simulation, the initial decrease in FCM
at the stop at $\Tstp=0.8\ (< \Tc)$ is clearly seen in
Fig.~\ref{fig:cl-T-cl-ht}(b),  and the decrease in FCM below
$\Tcsp$ during cooling can be recognized in our slowest-cooling
process with rate-10000 shown in the inset of Fig.~\ref{fig:fcm-zfcm_h}(b)
below. This FCM decrease below $\Tcsp$ is contrary to 
the FCM increase commonly observed in noninteracting magnetic
nanoparticle systems.\cite{SJTM} 

\begin{figure}
\begin{center}
\resizebox{0.32\textwidth}{!}{\includegraphics{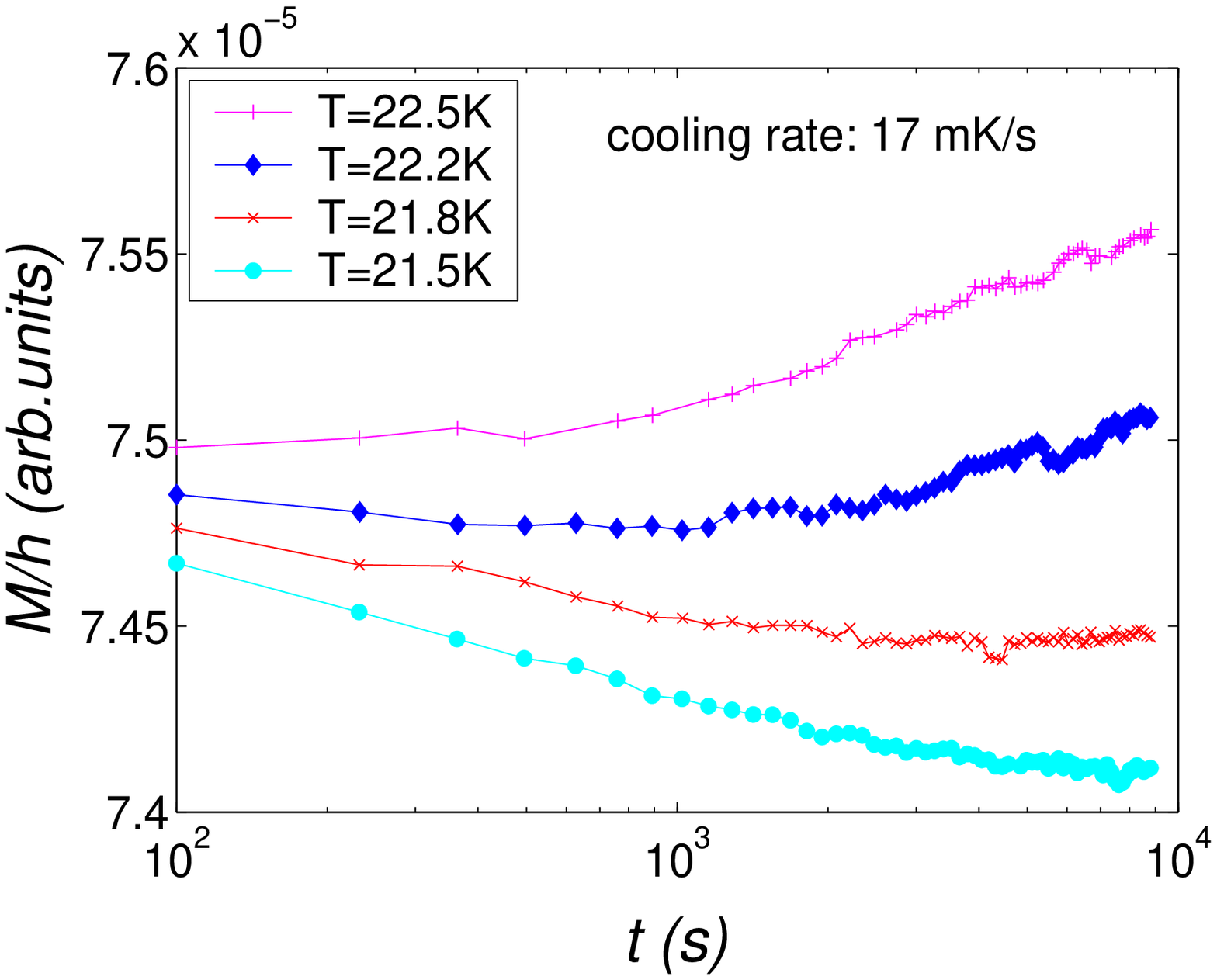}}
\end{center}
\vspace*{-3mm}
\caption{Time evolution of FCM during a stop of the cooling for \FeMnTiOII ($h=5$~Oe).
}
\label{fig:fcm-at-stop}
\end{figure}

During a stop at a temperature below $\Tcsp$, the SG short-range
order of size $\xi$ continues to increase and so the FCM decreases until
$\xi$ becomes comparable with $\lcrh$ given by Eq.~(\ref{eq:lcrh}). 
Then, $\xi$ no longer increases since the system is now in the
paramagnetic state. The magnetization makes a  turn and increases (due to the polarization of spin clusters) up to the
equilibrium paramagnetic value. This is the dynamical crossover scenario
proposed by TH. However, for $h$ of a few Oe, whose corresponding Zeeman
energy in \FeMnTiOII\ is about $10^{-4}\times\Tc$, $\lcrh$ is in general
much longer than $\xi^*$, and it will take an astronomical amount of
time for $\xi$ to reach $\lcrh$. An exception is expected at $\Tstp$
close to $\Tc$, at which $\lcrh$ can be small (Eq.~(\ref{eq:lcrh})) and
the growth rate of $\xi$ is the fastest. Actually, we experimentally
observed the expected FCM upturn behavior at such a stop as
shown in Fig.~\ref{fig:fcm-at-stop}. Similar behavior has been observed
for \FeMnTiO~\cite{T-Joensson}, while it is vaguely seen in the
simulation with rate-100 at $\Tstp=1.05$ (see inset of 
Fig.~\ref{fig:cl-T-cl-ht}(b)). 

Another finding observed in both Fig.~\ref{fig:FCM-ZFCM}(a) and 
Fig.~\ref{fig:fcm-zfcm_h}(b) is that the FHM is smaller than the FCM. This
can be simply regarded as the cumulative memory effect of the SG
short-range order.~\cite{JYMT} The SG order represented by $\xi$ 
increases at $T$ below $\Tcsp$ and so the depolarization 
increases but $\xi$ is still much shorter than $\lcrh$ when the system
is heated back to temperatures near $\Tcsp$. Also one can see that the
FCM at lowest temperatures is smaller, the slower the cooling
rate. In the simulation, this is observed in the cooling processes slower
than rate-333, as indicated in the inset of Fig.~\ref{fig:FCM-ZFCM}(b). In the
experimental results shown in Fig.~\ref{fig:FCM-ZFCM}(a), crossing
of the two FCM curves is observed. A possible interpretation for this is
that, for the slower cooling process, $\xi^*$ is longer and the
corresponding depolarization effect due to the SG stiffness energy below
$T^*$ is larger.  

Lastly, let us discuss the $h$-dependences of the FCM and ZFCM curves. For
the experimental results shown in Fig.~\ref{fig:fcm-zfcm_h}(a) one can
notice the following characteristics. As $h$ increases from 5~Oe, which we
have thus far investigated, $\Tirr$ approaches $T^*$. This implies
that a stronger field makes it easier for the ZFCM curve to merge with the
FCM curve in the paramagnetic phase. Under a sufficiently large $h$, on
the other hand, $\Tirr$ becomes lower than $T^*$, whose associated
cusp-like shape is much more rounded. The phenomena in this field range
coincide with those observed for \FeMnTiO\ by Aruga Katori and
Ito,~\cite{KatoriIto94} and are interpreted as the
dynamical crossover by TH; $\xi$ in the ZFCM process reaches a
relatively shorter $\lcrh$ at $T \approx \Tirr$ within the observation
time at each temperature. In the processes under $h$ of a few hundred
Oe, we see $\Tirr \approx T^*$. A rounded cusp-like shape in the FCM
curve in these processes can be regarded as a blocking phenomenon since
a difference between the FCM curves with different cooling rates is still 
noticeable below $T^*$ though it decreases with increasing $h$. 
As shown in Fig.~\ref{fig:fcm-zfcm_h}(b), $\Tirr$ obtained
by the simulation also decreases significantly as $h$ increases.

We have argued that the FCM behavior of Ising spin glasses observed in both
the experiment and simulation can be interpreted in a unified way by
a scenario based on the (extended) droplet picture, thereby we
have emphasized the viewpoint of `blocking' of the FCM, or `glassy
dynamics'. The scenario proposed in the present work is rather intuitive
and qualitative, but, at least at low temperatures under relatively
large $h$, it is consistent with the results obtained in our related
works,~\cite{oursLoren,TH} where we have 
argued that the experimental and simulation results, whose timescales
in units of microscopic spin-flip time differ by more than ten orders
of magnitude, can be understood in a unified way even 
semiquantitatively. Combined with our most recent work claiming the
dynamical breakdown of the SG state under relatively small 
$h$,~\cite{our-ac} the
present results strongly suggest that the slow dynamics in the FCM
processes we have observed under moderate $h$ is `glassy dynamics' far
from equilibrium.      

The viewpoint of `glassy dynamics' is a natural consequence of the 
marginal stability of the SG phase predicted by both the droplet theory and
the mean-field theory. It implies an infinitely wide distribution of
relaxation times, at least under $h=0$. In the 1980's, the peculiar FCM
phenomena discussed thus far had already been observed in various spin
glasses and were reviewed by Lundgren {\it et al}.~\cite{Lundgren} They
introduced a finite equilibration time, $t_{\rm eq}$, at all
temperatures; however, the above-mentioned marginal stability implies an
infinite $t_{\rm eq}$ at $T \le \Tc$ under $h=0$. According to the
droplet theory $t_{\rm eq}$ is expected to become finite under $h>0$ and
can be estimated using eq.(\ref{eq:lcrh}) combined with an appropriate
growth law for the SG short-range 
order.~\cite{Kisker-96,Marinari-growthLaw,ours1} For a
sufficiently small $h$, this $t_{\rm eq}$ becomes of an astronomical
order, and so the blocking of certain modes, spin clusters in the
present argument, is inevitable in real FCM measurements. Practically,
however, the FCM cusp is a good estimate of $\Tc$, i.e.,
 $T^* \simeq \Tc$,  if $h$ is small and the cooling rate is sufficiently
 slow. 
A detailed scenario of `glassy dynamics' for Heisenberg spin glasses is 
of importance, but it may become
more complicated due to the presence of two types of degrees of
freedom: continuous spin and (Ising-like) 
chirality~\cite{Petit,imakaw2004}. It is also of interest 
to analyze the dynamic transition of structural glasses on the
basis of the present scenario of `glassy dynamics'.

We thank H. Aruga Katori, A. Ito, K. Hukushima and P. Nordblad for
discussions and cooperation.  
P.E.J. acknowledges financial support
from the Japan Society for the Promotion of Science.
The present work is supported by NAREGI Nanoscience Project,
all from the Ministry  of Education, Culture, Sports, Science, and
Technology. The numerical simulations were performed using the
facilities at the Supercomputer Center, of the Institute for Solid State
Physics, at the University of Tokyo.


\begin{thebibliography}{99}
\bibitem{foot-Tstar}
An alternative definition of $\Tcsp$, more suitable for Fig.~\ref{fig:FCM-ZFCM}, 
is the temperature at which the FCM and FHM merge.
\end{thebibliography}
\end{document}